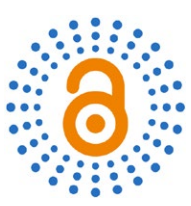

# Certain Amplified Genomic-DNA Fragments (AGFs) May Be Involved in Cell Cycle Progression and Chloroquine Is Found to Induce the Production of Cell-Cycle-Associated AGFs (CAGFs) in *Plasmodium falciparum*

**Gao-De Li**

Chinese Acupuncture Clinic, Liverpool, UK
Email: gaode_li@yahoo.co.uk

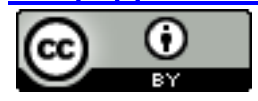

## Abstract

**It is well known that cyclins are a family of proteins that control cell-cycle progression by activating cyclin-dependent kinase. Based on our experimental results, we propose here a novel hypothesis that certain amplified genomic-DNA fragments (AGFs) may also be required for the cell cycle progression of eukaryotic cells and thus can be named as cell-cycle-associated AGFs (CAGFs). Like fluctuation in cyclin levels during cell cycle progression, these CAGFs are amplified and degraded at different points of the cell cycle. The functions of CAGFs are unknown, but we speculate that CAGFs may be involved in regulation of gene expression, genome protection, and formation of certain macromolecular complexes required for the dynamic genome architecture during cell cycle progression. Our experimental results also show that chloroquine induces the production of CAGFs in *Plasmodium falciparum*, suggesting that targeting cell cycle progression can be the primary mechanism of chloroquine's antimalarial, anticancer, and immunomodulatory actions.**

## 1. Introduction

Around 17 years ago, we performed an experiment in which both *Plasmodium falciparum* chloroquine (CQ)-resistant K1 isolate (K1 isolate) and *P. falciparum* CQ-sensitive HB3 isolate (HB3 isolate) were treated with CQ and their genomic DNA was isolated at different intervals and then subjected to PCR using an arbitrary primer. The results of the experiment showed that the PCR banding patterns displayed time dependent or cell cycle dependent changes in both control and CQ-treated groups [1]. Since we could not find a reasonable explanation for the PCR results and also because at that time our research interests were mainly focussed on finding out genetic differences between CQ-resistant and CQ-sensitive isolates, which eventually led to the discovery of a novel *P. falciparum* CQ resistance marker protein (Pfcrmp) gene [2] [3], we did not carry out any further investigation into the reason why the PCR banding patterns exhibited cell cycle dependent changes in both control and CQ-treated groups.

Recently, I read some articles about cyclins which were discovered in 1982 [4], and were very important proteins in controlling cell cycle progression by activating cyclin-dependent kinase [5] [6]. The fluctuation in cyclin levels during cell cycle progression reminds me of rethinking about the cell cycle dependent changes of PCR banding patterns described above. Is it possible that certain genomic-DNA fragments are amplified and degraded at different points of the cell cycle? The answer should be yes because only the fluctuation of these genomic-DNA fragments during cell cycle progression can be used to reasonably explain the PCR results and therefore a novel hypothesis is proposed in this paper.

## 2. The Hypothesis

Certain genomic-DNA fragments may be amplified at different points of eukaryotic cell cycle. Instead of being incorporated into the cell's genome or chromosomes, the amplified genomic-DNA fragments (AGFs) are released into the nucleoplasm. These "extragenomic" AGFs are required for cell cycle progression and thus could be named as cell-cycle-associated AGFs (CAGFs). Like fluctuation in cyclin levels during cell cycle progression, the CAGFs are amplified and degraded at different points of the cell cycle.

## 3. Experimental Evidence to Support the Hypothesis

CQ has been the mainstay of malaria chemotherapy for many decades and is now widely used in treatment of other conditions including being used as an adjunct in anticancer therapy [7] [8], but its mechanism of action remains unresolved. One of hypotheses about the mechanism of CQ's action is that CQ may exert its action on DNA either through intercalating into DNA or by inducing apoptosis [9]-[11].

In order to explore if CQ can cause genomic-DNA damage, around 17 years ago we performed the following experiment that no one else seems to have performed.

### 3.1. Experimental Methods

#### 3.1.1. Genomic-DNA Isolation

Malaria parasites (K1 and HB3 isolates) were synchronised at the ring stage which is the earliest stage during the intra-erythrocytic cycle of malaria parasite [12] and exposed to CQ at various concentrations. CQ treated and control parasites were cultured under our laboratory routine conditions and harvested at 2 h, 6 h, 12 h, and 24 h after CQ treatment. The harvested parasites were incubated with lysis buffer at 50˚C overnight. The lysates were extracted twice with phenol/chloroform and the genomic DNA was precipitated with 0.3 M sodium acetate and 2 volumes of ethanol. After RNase I treatment, the quantity of DNA was measured using a Spectrophotometer.

#### 3.1.2. Arbitrarily-Primed PCR

Only one arbitrary primer named as AP1 (5'-GAATTCGCGGCCGCAGGAAT-3') was used in the PCR. Genomic DNA 10 ng, 20 pmol AP1 and 1.5 u Taq DNA polymerase (Bioline) were added to 50 µl of PCR reaction mixture. Forty PCR cycles were performed, each consisting of denaturation at 94˚C for 30 seconds, primer annealing at 60˚C for 30 seconds and primer extension at 72˚C for 1 minute. The denaturation step in the first cycle was extended to 4 minutes. The PCR products were separated by agarose gel electrophoresis, stained with ethidium bromide and visualised under ultraviolet light.

### 3.2. Results and Discussion

Both K1 and HB3 isolates were treated with CQ at their respective $IC_{10}$, $IC_{50}$, and $IC_{99}$ concentrations. Genomic DNA was isolated at intervals of 2 h, 6 h, 12 h, and 24 h after CQ treatment, and subject to arbitrarily-primed PCR using an arbitrary primer named as AP1. The results showed that there were two distinct types of PCR banding patterns. Bands that remained unchanged throughout the sampling interval (2 h to 24 h) are referred to as stable bands. Five stable bands were found in PCR products of genomic-DNA samples from both K1 and HB3 isolates. Bands that were not produced consistently throughout the sampling period are referred to as unstable bands (UB). Two such unstable bands named as UB1 and UB2, respectively were identified in PCR products of genomic-DNA samples from both K1 and HB3 isolates. The appearance of UB1 and UB2 displayed time dependent or cell cycle dependent changes in both control and CQ-treated groups (**Figure 1**).

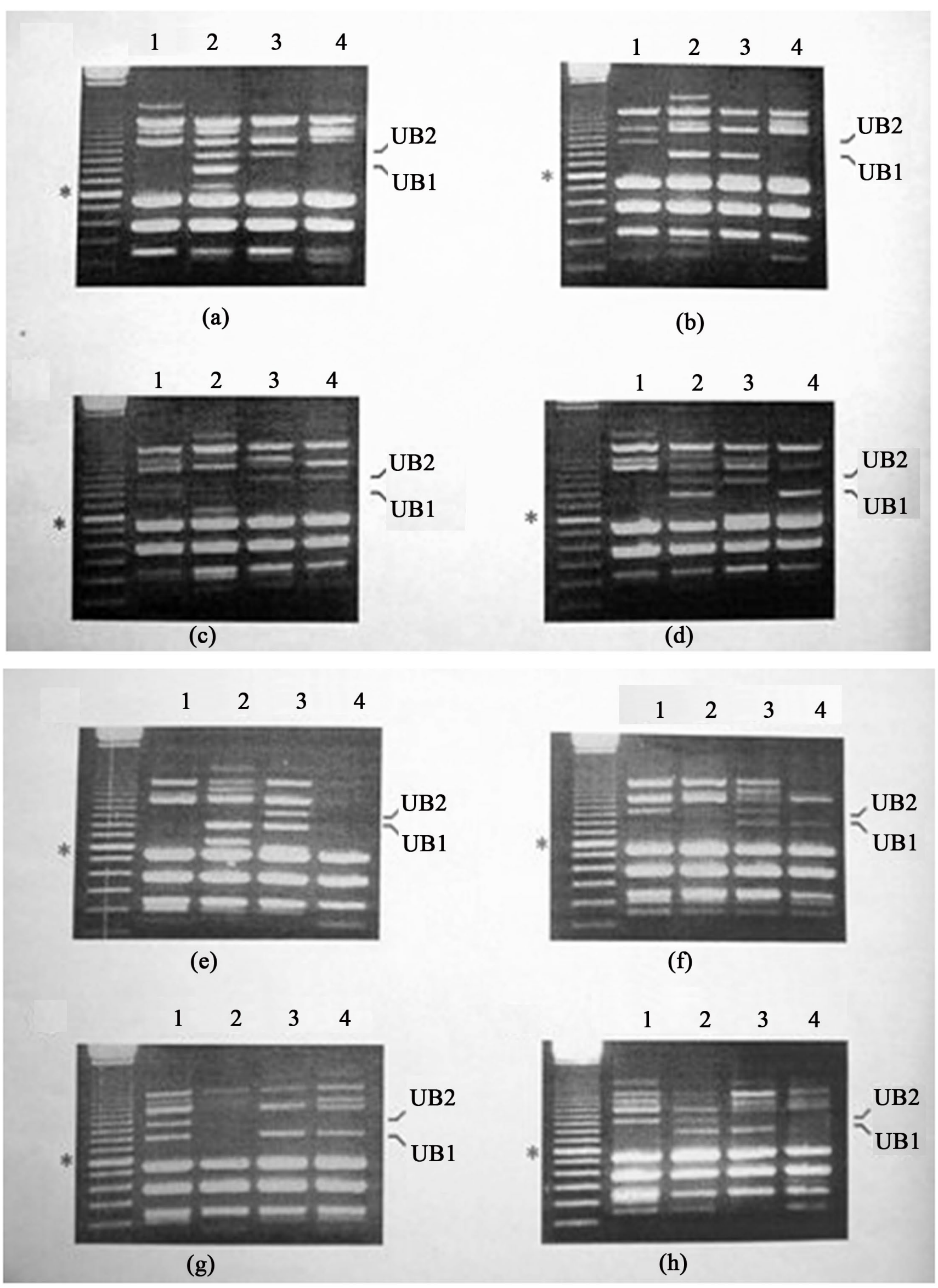

**Figure 1**. PCR banding patterns from the CQ-resistant K1 isolate (a-d) and CQ-sensitive HB3 isolate (e-h) of *P. falciparum*. (a) and (e) = 2 h post CQ treatment; (b) and (f) = 6 h post CQ treatment; (c) and (g) = 12 h post CQ treatment; (d) and (h) = 24 h post CQ treatment. Lanes 1 = control groups, 2 = CQ $IC_{10}$, 3 = CQ $IC_{50}$, 4 = CQ $IC_{99}$. Asterisks indicate 800 bp band in the 100-bp-DNA ladder lane.

We also cloned and sequenced the UB1 band. The results showed that it had AP1 complimentary sequences at both ends, indicating that the UB1 was produced by one single primer annealed at two counter-directional sites of a genomic-DNA fragment. The sequence between the two AP 1 sites showed typical codon usage for *P. falciparum* [13]. Blast search against amino acid databases revealed that the sequence shared 80% similarity with a gene of unknown function from *Escherichia coli* (P32667). Based on the full sequence of this fragment two internal specific primers (SP1: 5'-AATAAGATGTGCTGTTTAACT-3' and SP2: 5'-GTTGTGCTCACACTTATCT-3') adjacent to the AP1 sites were designed and synthesized. The expected PCR band produced with SP1 and SP2 is 68 bp smaller than UB1 band and thus could be named as UB1-related band. Using SP1 and SP2, 2 h-isolated DNA samples from both K1 and HB3 isolates were rechecked by PCR. The results showed that only one single UB1-related band was found in all samples, but the PCR bands of the CQ-treated group were much brighter and bigger than those of the control group (data not shown).

In theory the genomic DNA from control groups should remain unchanged not only in nature but also in quantity throughout the 24 h sampling period in both K1 and HB3 isolates because DNA synthesis in *P. falciparum* takes place 29 h after the intra-erythrocytic cycle started [14]. If genomic DNA remains unchanged the PCR banding patterns in the control groups should remain the same from 2 h to 24 h. As for the CQ-treated groups, if the genomic DNA of malaria parasite could be damaged by CQ treatment the number of PCR bands should be smaller in the CQ-treated groups than in the control groups because the genomic-DNA damage means that the quality and quantity of template for AP1 will be affected. Surprisingly, the PCR results described above were totally different from what we anticipated. Since at that time we could not figure out why the UB1 and UB2 in both control and CQ-treated groups exhibited cell cycle dependent changes, we didn't carry out any further investigation, nor did we publish the results although the PCR results were highly reproducible.

Inspired by cyclin's discovery, we now think that the PCR results could be reasonably explained by the hypothesis we proposed above. In order to easily observe the cell cycle dependent changes in UB1 and UB2, we summarized the appearance (marked as +) and disappearance (marked as −) of both UB1 and UB2 in two tables (**Table 1**, **Table 2**).

The PCR templates for both UB1 and UB2 could be considered to represent two different unknown-sized CAGFs. In the arbitrarily-primed PCR analysis of 2h-isolated samples from both K1 and HB3 isolates, neither UB1 band nor UB2 band was found in the PCR products of the control groups, however when two specific primers (SP1 and SP2) were used in the PCR, UB1-related band (equivalent to UB1) was found in the PCR product of the control groups (No doubt UB2-related band could also be found in the PCR product of the control groups if two internal specific primers were used in the PCR). The reason for this difference is that for arbitrarily-primed PCR analysis the number of templates for UB1 in 2h-isolated genomic DNA from the control groups may not be enough because arbitrary primer due to its mispaired annealing usually is much less efficient than specific primers in PCR analysis and therefore to produce a visible UB1 band in PCR products more genomic-DNA templates for UB1 are required for arbitrarily-primed PCR compared to the PCR using specific primers. In a word, the reason why UB1 and UB2 displayed cell cycle dependent changes in the control groups of both K1 and HB3 isolates is because the number of genomic-DNA templates for them was increased and decreased at different points of the cell cycle, indicating that the related genomic-DNA fragments, *i.e.* CAGFs were amplified and degraded at different points of the cell cycle. Since CAGFs are easily amplified and degraded during cell cycle progression, they are certainly not incorporated into the genome, but released from the genome into the nucleoplasm.

**Table 1.** Fluctuation of UB 1 and UB2 bands in the PCR products from the genomic-DNA samples of K1 isolate.

| | | 2 h | 6 h | 12 h | 24 h | Total + |
|---|---|---|---|---|---|---|
| **Control** | UB1 | − | − | + | − | 3 |
| | UB2 | − | + | + | − | |
| **CQ** | UB1 | + | + | − | + | 5 |
| $IC_{10\ (25nM)}$ | UB2 | + | − | − | + | |
| **CQ** | UB1 | − | + | − | − | 4 |
| $IC_{50\ (150nM)}$ | UB2 | + | − | + | + | |
| **CQ** | UB1 | − | − | − | + | 2 |
| $IC_{99\ (500nM)}$ | UB2 | − | − | + | − | |

**Table 2.** Fluctuation of UB 1 and UB2 bands in the PCR products from the genomic-DNA samples of HB3 isolate.

| | | 2 h | 6 h | 12 h | 24 h | Total + |
|---|---|---|---|---|---|---|
| **Control** | UB1 | − | − | + | − | 4 |
| | UB2 | − | + | + | + | |
| **CQ** | UB1 | + | − | − | + | 3 |
| $IC_{10\ (5nM)}$ | UB2 | − | − | − | + | |
| **CQ** | UB1 | + | + | + | + | 6 |
| $IC_{50\ (25nM)}$ | UB2 | + | + | − | − | |
| **CQ** | UB1 | − | + | + | − | 2 |
| $IC_{99\ (250nM)}$ | UB2 | − | − | − | − | |

As mentioned above, when the 2 h-isolated genomic-DNA samples were rechecked with PCR using SP1 and SP2, the UB1-related bands in the groups treated with CQ at 25 nM concentration were found to be much brighter and bigger than those in the control groups. A reasonable explanation for these differences is that the number of genomic-DNA templates for the UB1 was increased in the CQ-treated groups although at that time we didn't perform quantitative PCR analysis.

The reason why both UB1 and UB2 appeared in the PCR products of 2h-isolated CQ-treated samples (except $IC_{99}$ -treated samples) from both K1 and HB3 isolates is because CQ could induce the production of UB1 and UB2-related CAGFs in malaria parasite. Based on the PCR results of 2h-isolated samples (**Figure 1(a)** lane 2 and **Figure 1(e)** lane 3) the best CQ concentration for inducing both UB1 and UB2-related CAGFs was 25nM in both K1 ($IC_{10}$ = 25 nM), and HB3 ($IC_{50}$ = 25 nM) isolates. Overall, the $IC_{99}$ was the poorest CQ concentration for inducing the UB1 or UB2-related CAGFs in both K1 and HB3 isolates. The same conclusion can be drawn from checking the total + number in the CQ-treated groups (**Table 1**, **Table 2**). CQ induces the production of CAGFs in *P. falciparum*, which could greatly affect intraerythrocytic cycle progression of malaria parasite， suggesting that targeting cell cycle progression could be the primary mechanism of CQ's antimalarial, anticancer and immunomodulatory actions [7]. The mechanism by which CQ induces the production of CAGFs in malaria parasite is unknown, presumably through activation of an unknown signalling pathway.

According to our experimental results, CAGFs were synthesized or amplified before cell cycle progression into DNA synthesis phase (S phase) of *P. falciparum*, which indicates that an unknown DNA synthesis mechanism might exist outside S phase of the cell cycle. Like inducible mRNA transcripts, CAGFs are genome-released, drug-inducible and easily degraded during the cell cycle progression, which enables us to speculate that CAGFs might be single-stranded DNA fragments that are synthesized through the process of "DNA-to-DNA transcription", which is similar to DNA-to-RNA transcription, but using DNA polymerase instead of RNA polymerase. In other words, outside the S phase of the cell cycle, DNA replication scattered-increase could be thought of as DNA to DNA transcription.

At present, we don't know how many different CAGFs are involved in the cell cycle progression of eukaryotic cells, and how many different CAGFs could be induced by CQ treatment. The primer, AP1 used in our experiment can only detect UB1 and UB2-related CAGFs, for detecting other CQ-induced CAGFs more different arbitrary primers should be designed and tested. The functions of CAGFs are unknown, but we speculate that CAGFs might be involved in regulation of gene expression, and formation of certain macromolecular complexes required for the dynamic genome architecture during cell cycle progression [15]-[18]. Furthermore, since the CAGFs are not chromosomal DNA that is well protected by proteins, once they are amplified and released into the nucleoplasm, they could easily be exposed to DNA-damaging agents that enter the nucleus, absorbing the attack effects, and thus might function as bodyguards to protect the genome from damage.

That CQ induces the production of CAGFs in malaria parasite could also be interpreted as a genome defence response against DNA-damaging agents. But at the same time more CAGFs produced in the nucleus might cause chaos in the regulation of cell cycle progression, which may contribute to understanding of the mechanism of CQ's diverse therapeutic actions.

Many studies have shown that CQ can cause cell cycle arrest in various cancer cells [19]-[21]. If the cell cycle arrest is mediated by CQ-induced CAGFs that might intervene in gene expression and regulation during cell cycle progression, the CAGFs could be deemed as "brake pads" that slow down or stop the cell cycle progression. Possibly, cancers might arise from the cells in which the brake failure has occurred.

## 4. Conclusion

The hypothesis proposed in this paper is based on our preliminary experimental results obtained from *P. falciparum*. Whether it can stand up to further testing in other eukaryotic cells, more investigations are needed. In order to provide convenience to those who want to repeat our experiment, we have presented the DNA sequences of two specific primers (SP1 and SP2) in this paper. Hopefully, through further studies using a wide range of methods, the hypothesis may eventually be incorporated into the current theories of cell cycle regulation and enrich our knowledge of DNA functions.

## Acknowledgements

The experiments were designed and performed by the author around late 1998 to early 1999 in University of Liverpool, UK and supported by a grant from Wellcome Trust. At that time the author was a Chinese scientist from Second Military Medical University, Shanghai, China.